\documentclass[
  aps,
  preprint,
  amsmath,
  amssymb,longbibliography,
  nofootinbib
]{revtex4-1}

\usepackage[T1]{fontenc}
\usepackage[utf8]{inputenc}
\usepackage{bm}
\usepackage{graphicx}
\usepackage{hyperref}
\usepackage{slashed}

\hypersetup{
  colorlinks=true,
  linkcolor=blue,
  citecolor=magenta,
  urlcolor=blue
}

\DeclareMathOperator{\Tr}{Tr}

\begin{document}

\title{Quantum Gravity Beyond the Bulk}

\author{Jorge Gamboa}
\affiliation{Departamento de Física, Universidad de Santiago de Chile, Santiago, Chile }
\email{jorge.gamboa@usach.cl (corresponding author)}

\author{Natalia Tapia-Arellano}
\affiliation{Department of Physics and Astronomy, Agnes Scott College, Decatur, GA. 30030, USA}
\email{narellano@agnesscott.edu}

\date{\today}


\begin{abstract}
We propose an infrared and asymptotic formulation of quantum gravity
adapted to external observers. In an asymptotic proper-time gauge,
the physical generator of evolution reduces to the Regge--Teitelboim
boundary charge, so that quantum dynamics is governed by infrared
gravitational configurations at spatial infinity. In the weak-field
regime this leads naturally to a Born--Oppenheimer separation between
slow asymptotic data and fast bulk fluctuations. Integrating out the
latter induces a Berry connection on the space of admissible
configurations, whose holonomy characterizes infrared-dressed
gravitational states. Observable evolution is then governed by an
infrared effective Hamiltonian obtained after integrating out fast
bulk fluctuations, with the geometric (Berry) contribution entering
as part of the resulting evolution operator on the space of
asymptotic configurations. Tracing over unresolvable degrees of
freedom induces a reduced density matrix whose entropy is dominated
by global infrared holonomy sectors rather than local bulk dynamics.
\end{abstract}


\maketitle


{\bf Introduction}: The standard formulation of quantum mechanics and quantum field theory is inherently local \cite{Glimm:1987ylb}:
physical observables are associated with spacetime regions, time is defined locally, and the observer may be idealized as external to the system \cite{Wheeler:1984dy}. While this framework is fully adequate for non--gravitational interactions, it becomes conceptually strained once gravity is taken into account \cite{Isham:1992ms}. Gravitational dynamics determines the global causal structure of spacetime, and fundamental observables such as energy, momentum, and angular
momentum acquire an unambiguous meaning only at infinity \cite{AshtekarStreubel1981,BrownYork1993,ReggeTeitelboim1974,BrownYork1993}. The presence
of horizons further enforces an operational loss of information,
compelling an external observer to describe the system by reduced
quantum states.

These features suggest that the quantum gravitational information
accessible to asymptotic observers is not primarily encoded in local bulk operators, but rather in asymptotic and infrared structures
\cite{Weinberg:1965nx,Bondi:1962px,Sachs:1962zza,Strominger:2017zoo}.
In this sense, quantum gravity is naturally closer in spirit to the infrared sector of gauge theories, where physical states are not Fock states but infrared--dressed configurations characterized by soft modes, geometric phases, and entanglement
\cite{Chung:1965zza,Kibble:1968lka,Kulish:1970ut,Weinberg:1965nx,Gamboa:2025dry,Gamboa:2025fcn,
Gamboa:2025qjr}.
In gauge theories, this structure can be described in a controlled way through an adiabatic separation between slow (infrared) and fast
(ultraviolet) degrees of freedom, leading to functional Berry connections and holonomies on configuration space \cite{Berry:1984jv}.

In this essay, we adopt this infrared and asymptotic perspective and apply it directly to gravity. Rather than pursuing a full quantization of the bulk or addressing the Wheeler--DeWitt equation
\cite{DeWitt:1967, Klauder:1972je} in its most general form, we focus on the asymptotic weak--field regime relevant to an external observer.
Within this framework, the Hamiltonian constraint acts as a
compatibility condition governing transport in the space of asymptotic configurations, while physical evolution is generated by boundary charges defined at infinity \cite{ReggeTeitelboim1974}.

Using a canonical framework and an adiabatic Born--Oppenheimer--type separation of gravitational degrees of freedom, we show that infrared gravitational states are naturally characterized by functional holonomies associated with Berry \cite{Berry:1984jv}
connections on the space of admissible asymptotic data, rather than by local bulk excitations.

This viewpoint leads to a geometric interpretation of infrared gravitational information, in which asymptotic symmetries, conserved charges, and superselection sectors emerge as effective descriptors of dressed states. 

As an application, we discuss black holes as infrared laboratories, where the limited resolving power of the asymptotic charge algebra effectively induces an operational coarse--graining of gravitational states. Tracing over degrees of freedom that are not independently resolvable within the asymptotic algebra leads to a mixed state, whose entropy receives a natural contribution from inequivalent infrared holonomy sectors. In the extremal case this mixture is
not thermal in origin, but reflects the existence of infrared superselection sectors that cannot be distinguished by boundary observables.

From this perspective, quantum gravity appears as an effective, geometric theory of the asymptotic sector, with entropy dominated by global infrared structures rather than local bulk dynamics.

\medskip
{\bf Developing the Idea}: We now turn to gravity and adopt a canonical description in which the notion of time is fixed by a choice of foliation, as appropriate for an external (asymptotic) observer. Concretely, we work, in the asymptotic region, in a convenient gauge by imposing
\cite{Nambu:1950rs,Feynman:1949hz,Schwinger:1951xk,Teitelboim:1983fk}
\begin{equation}
\dot N_\perp = 0, \qquad N_i = 0,
\end{equation}
so that the shift vanishes and the lapse is taken to be constant, which we set to unity for simplicity. This asymptotic gauge choice preserves the constraint structure while allowing the evolution parameter to be identified with the proper time measured at spatial infinity, in analogy with the proper--time gauge of the relativistic particle.

Importantly, this choice does not remove the evolution parameter itself; rather, it removes the \emph{local arbitrariness} in its
definition and makes explicit that the bulk gravitational
Hamiltonian remains a constraint.

As a consequence, the flow generated by the bulk Hamiltonian
constraint is pure gauge at the microscopic level: it moves states along gauge orbits in phase space and therefore does not define an independent physical time evolution in the bulk, but merely parametrizes different descriptions of the same physical configuration. This does not preclude an intrinsic asymptotic
evolution, which arises only after supplementing the bulk
constraints with the appropriate Regge--Teitelboim surface term
\cite{ReggeTeitelboim1974}.

In generally covariant theories the canonical Hamiltonian takes
the form
\begin{equation}
H_{\rm can}[N_\perp,N^i]
=
\int_\Sigma d^3x\;
(N_\perp\mathcal H_\perp + N^i\mathcal H_i),
\end{equation}
where $\mathcal H_\perp$ and $\mathcal H_i$ denote the Hamiltonian and momentum constraints, respectively.

However, the canonical generator $H_{\rm can}$ is not functionally
differentiable on the space of asymptotically flat configurations. Its variation contains a non--vanishing surface contribution of the form
\begin{equation}
\delta H_{\rm can}
=
(\text{bulk})
+
\oint_{\partial\Sigma}
\Theta[h_{ij},\delta h_{ij}],
\end{equation}
which prevents $H_{\rm can}$ from generating well--defined canonical transformations. As first shown by Regge and Teitelboim
\cite{ReggeTeitelboim1974}, this obstruction can be removed by
introducing a boundary functional $Q_{RT}$ such that
\begin{equation}
\delta Q_{RT}
=
-
\oint_{\partial\Sigma}
\Theta[h_{ij},\delta h_{ij}].
\end{equation}
The total differentiable Hamiltonian is then defined as
\begin{equation}
H_{\rm tot}[N_\perp,N^i]
=
H_{\rm can}[N_\perp,N^i]
+
Q_{RT}[N_\perp,N^i],
\end{equation}
whose variation is purely bulk and therefore generates
well--defined asymptotic time translations.

On the constraint surface $\mathcal H_\perp\approx 0$ and
$\mathcal H_i\approx 0$, or after projecting onto physical states, the bulk contribution vanishes and the generator of evolution relevant to an asymptotic observer reduces entirely to the
surface term,
\begin{equation}
H_{\rm phys}=Q_{RT}.
\end{equation}

Consequently, any notion of quantum evolution accessible to an external observer must be formulated in terms of asymptotic data and boundary generators, rather than local bulk operators.

Fixing an asymptotic proper--time gauge ($N_\perp=1$, $N_i=0$)
selects a preferred foliation at spatial infinity and defines
an intrinsic proper-time ($s$) associated with the
asymptotic observer. In this gauge, the quantum evolution of
infrared configurations may be written in the Schr\"odinger--type form
\begin{equation}
i\frac{\partial}{\partial s}
\Psi[h_1,h_2;s]
=
\widehat{Q}_{RT}
\Psi[h_1,h_2;s],
\label{eq:proper_time_Sch}
\end{equation}
which describes the evolution of asymptotically resolvable
infrared configurations.

In the asymptotic weak--field region we expand the spatial
metric as
\begin{equation}
g_{ij}=\delta_{ij}+h_{ij},\qquad |h_{ij}|\ll 1,
\end{equation}
and organize the Wheeler--DeWitt equation as a perturbative
expansion in powers of $h_{ij}$. 

\medskip

In this work we adopt the operational standpoint that any physically
meaningful formulation of quantum gravity must be defined with respect
to an external asymptotic observer. From this perspective, genuinely
quantum information cannot be encoded in local bulk operators, but must
instead reside in infrared degrees of freedom associated with the
asymptotic structure of spacetime.

This suggests that quantum gravity should be formulated as an effective
infrared theory in which slow gravitational modes are treated dynamically,
while fast ultraviolet modes play the role of an environment inducing
geometric (Berry-type) phases on the space of asymptotic configurations.

\medskip

The DeWitt supermetric
admits a corresponding weak--field expansion,
\begin{equation}
G_{ijkl}(g)
=
G^{(0)}_{ijkl}
+
G^{(1)}_{ijkl}[h]
+
\mathcal O(h^2).
\end{equation}

To leading order in the weak--field expansion, this structure allows one
to consistently replace the full DeWitt supermetric by its background
value, yielding

\begin{equation}
G^{(0)}_{ijkl}
=
\frac12\Big(\delta_{ik}\delta_{jl}
+\delta_{il}\delta_{jk}
-\delta_{ij}\delta_{kl}\Big).
\end{equation}

At leading order the supermetric may consistently be evaluated
on the flat background,
\begin{equation}
G_{ijkl}(g)\;\longrightarrow\;G^{(0)}_{ijkl}(\delta),
\end{equation}
with all metric--dependent corrections relegated to
higher--order terms. This choice ensures that the leading
Schr\"odinger--type operator is purely kinematical, while
genuine dynamical and geometric effects enter through the
structure of the wave functional itself and through the
boundary generator.

From the asymptotic viewpoint adopted here, the physically
relevant gravitational data accessible to an external
observer are encoded in boundary charges such as $Q_{RT}$,
which depend only on the infrared behaviour of the metric
at spatial infinity. This suggests organizing the theory
in a Born--Oppenheimer (adiabatic) manner, separating slow collective configurations that determine the asymptotic geometry from fast bulk fluctuations that are not
independently resolvable at infinity.

Accordingly, we decompose the asymptotic metric perturbation as
\begin{equation}
h_{ij}=h_{1\,ij}+h_{2\,ij},
\end{equation}
where $h_1$ parametrizes slow infrared configurations
associated with admissible asymptotic data, while $h_2$
denotes fast fluctuations encoding ultraviolet information
about the bulk. The distinction between slow and fast modes should not be understood in terms of a sharp momentum
cutoff, but rather as an adiabatic separation defined on
the space of asymptotic configurations, in direct analogy
with the Born--Oppenheimer construction in gauge theories.

Writing the wave functional as
\begin{equation}
\Psi[h_1,h_2;s]=\chi[h_1;s]\;\phi_0[h_2;h_1],
\end{equation}
where $\phi_0[h_2;h_1]$ denotes the instantaneous reference state of the fast sector at fixed $h_1$ within the asymptotic weak--field regime, and projecting Eq.~(\ref{eq:proper_time_Sch}) onto this state, one obtains an effective evolution equation for the slow wave functional $\chi[h_1;s]$. The dependence of $\phi_0$ on the slow variables implies that functional derivatives acting on $\Psi$ generate additional geometric terms, which combine into a Berry connection on the space of asymptotic configurations,
\begin{equation}
\mathcal A^{ij}(x;h_1)
:=
i\,
\big\langle 0;h_1\big|
\frac{\delta}{\delta h_{1\,ij}(x)}
\big|0;h_1\big\rangle .
\end{equation}

As a result, the adiabatic response of the slow sector is encoded in
covariant functional derivatives,
\begin{equation}
-i\frac{\delta}{\delta h_{1\,ij}(x)}
\;\longrightarrow\;
-i\frac{\delta}{\delta h_{1\,ij}(x)}
-\mathcal A^{ij}(x;h_1),
\end{equation}
reflecting parallel transport in the space of slow configurations.
Evaluated along an adiabatic trajectory $\Gamma:s\mapsto h_1(s)$,
this induces the transport law
\begin{equation}
\left(
\frac{d}{ds}
+
i\int d^3x\,
\dot h_{1\,ij}(x,s)\,
\mathcal A^{ij}(x;h_1)
\right)
\chi[h_1(s)]
=0,
\label{eq:adiabatic_transport}
\end{equation}
whose solution is the functional holonomy
\begin{equation}
\chi[h_1(s)]
=
\mathcal P\,
\exp\!\left(
-\,i\int_{\Gamma}
d h_{1\,ij}(x)\,
\mathcal A^{ij}(x;h_1)
\right)
\chi[h_1(s_0)] .
\label{eq:parallel_transport_eff}
\end{equation}

However, physical evolution for an asymptotic observer arises
only after projecting the Schr\"odinger--type equation
\eqref{eq:proper_time_Sch} onto the instantaneous fast
reference state $\phi_0[h_2;h_1]$. This yields an effective
infrared dynamics for the slow wave functional,
\begin{equation}
i\frac{\partial}{\partial s}
\chi[h_1;s]
=
\widehat H_{\rm eff}[h_1]\,
\chi[h_1;s],
\end{equation}
where
\begin{equation}
\widehat H_{\rm eff}
=
\widehat Q_{RT}
+
\widehat V_{\rm BH}
+
\widehat{\mathcal A}.
\end{equation}

From the viewpoint of an external observer, the infrared
gravitational sector is therefore described by an effective
quantum field theory on an asymptotically flat background,
with evolution generated by a self--adjoint Hamiltonian
constructed from the boundary charge and geometric
Born--Huang (BH) corrections.
The effective evolution of the slow wave functional
$\chi[h_1;s]$ is therefore governed by an infrared
Hamiltonian containing both dynamical and geometric
(Born--Huang and Berry) contributions. The functional
holonomy associated with the Berry connection enters
as the geometric component of the resulting evolution
operator on the space of asymptotic configurations.

{\bf Interpretation and physical content.}
The above construction shows that, once infrared dressing and a Born--Oppenheimer separation between slow and fast gravitational degrees of freedom are implemented, the Wheeler--DeWitt constraint admits a natural interpretation as a covariant transport condition on the space of admissible infrared configurations accessible to an external observer.

As discussed in the previous section, the bulk Hamiltonian generates pure gauge transformations and does not correspond to physical time evolution. Observable evolution arises only after supplementing the constraints with the appropriate Regge--Teitelboim surface term, whose quantized generator defines the asymptotic Hamiltonian $H_{\rm phys}=Q_{RT}$.

In this framework, the Wheeler--DeWitt equation should therefore
be viewed not as a microscopic dynamical equation in the bulk, but as a compatibility condition governing parallel transport in the space of slow configurations after projection onto the physical Hilbert space. The functional Berry connection encodes the geometric response of the fast gravitational sector to adiabatic changes of the asymptotic data, while its holonomy characterizes
physically inequivalent infrared--dressed gravitational states.

The resulting transport equation does not replace dynamics, but
represents its effective adiabatic form after the fast degrees of freedom have been integrated out. Physical evolution is generated by the boundary charge and may be formulated relationally as
transport along adiabatic trajectories in configuration space, or equivalently as evolution with respect to an intrinsic asymptotic parameter such as proper time.

When the slow parameter is identified with an RG scale, the
covariant transport law induced by the Wheeler--DeWitt constraint
becomes formally analogous, within the adiabatic regime, to a
renormalization--group equation written as a parallel--transport
condition, with the Berry connection playing the role of an
effective RG connection on the space of admissible configurations.

This formulation also clarifies the role of asymptotic symmetries
and conserved charges. Rather than acting as fundamental operators on a bare Hilbert space, they emerge as effective descriptors of the holonomy structure of infrared--dressed states. Distinct superselection sectors correspond to inequivalent holonomy classes in configuration space, while local bulk excitations enter only
indirectly through their influence on the geometric structure.

In this sense, quantum gravity appears as an intrinsically
infrared theory of dressed states, whose physical content is
encoded in geometric phases, functional holonomies, and
entanglement, rather than in local bulk observables. The
Born--Oppenheimer approximation provides a controlled and
physically transparent framework in which this structure becomes
manifest, placing gravity on the same geometric footing as the
infrared sector of gauge theories.

{\bf Global dressed state and reduced density matrix.}
Let $\Psi[h]$ denote the infrared--dressed gravitational state defined in the canonical framework, for instance through the projector
construction and the Born--Oppenheimer dressing ansatz introduced above. The exterior density matrix is defined by tracing over degrees of freedom that are not accessible to an asymptotic observer,
\begin{equation}
\rho_{\rm ext}
=
\Tr_{\rm int}\,|\Psi\rangle\langle\Psi|,
\qquad
S_{\rm vN}
=
-\Tr_{\rm ext}
\bigl(
\rho_{\rm ext}\ln\rho_{\rm ext}
\bigr).
\label{eq:SvN_def}
\end{equation}
Operationally, the trace $\Tr_{\rm int}$ reflects the fact that only observables associated with the asymptotic algebra generated by the Regge--Teitelboim charge $Q_{RT}$ are physically resolvable at infinity. Degrees of freedom that do not belong to this algebra, including fast bulk fluctuations or complementary infrared sectors, cannot be independently measured by an external observer and must
therefore be traced over in the effective infrared description.

{\bf Geometry and infrared structure in the extremal
BTZ case.}
For extremal BTZ black holes the near--horizon geometry differs qualitatively from the non--extremal Schwarzschild case. The
surface gravity vanishes, $\kappa=0$, and the Euclidean continuation does not impose any periodicity in the time coordinate. As a result, the standard Rindler reduction and the associated Hawking thermality are absent: the extremal geometry does not define a local thermal density matrix for exterior degrees of freedom.

This absence of thermality reflects a deeper structural property of three--dimensional gravity. Since there are no local propagating bulk degrees of freedom, the infrared sector is not governed by local
horizon physics but by global and asymptotic data. In particular, the physical phase space is entirely captured by the boundary degrees of
freedom associated with the asymptotic symmetry algebra \cite{BrownHenneaux1986}. For extremal BTZ black holes, the condition $J=M\ell$ freezes one
chiral sector, and the infrared dynamics reduces to a single Virasoro coadjoint orbit \cite{BrownHenneaux1986}.

In this setting, the reduced infrared phase space may be identified with this coadjoint orbit, which naturally plays the role of the slow configuration space in the adiabatic framework. Integrating out fast
bulk fluctuations induces a Berry connection on this orbit, whose holonomies classify inequivalent infrared--dressed gravitational states. Infrared degrees of freedom are therefore characterized not by local excitations near the horizon, but by geometric phases acquired under adiabatic transport along nontrivial loops in the space of boundary configurations.

{\bf Entropy and geometric phases.}
In the extremal BTZ case \cite{BTZ1992}, entropy admits a natural interpretation in terms of the loss of information associated with superselection sectors that are not distinguishable by the asymptotic
charge algebra generated by $Q_{RT}$.

Infrared dressing organizes the Hilbert space into inequivalent
holonomy sectors that cannot be related by local bulk operations.
Tracing over fast or complementary infrared degrees of freedom therefore induces a mixed state over these sectors. The resulting von Neumann entropy $S_{\rm IR}$ measures the topological entanglement
associated with the corresponding holonomy classes, rather than local field fluctuations.

From this viewpoint, extremal three--dimensional black holes realize a regime in which the infrared limit of quantum gravity is governed by geometric phases and topological data rather than by local thermality.

In this way, quantum gravity emerges not as a quantization of bulk
spacetime geometry, but as an infrared geometric theory governing
the parallel transport of asymptotic configurations. Observable
quantum gravitational effects are therefore expected to arise not
from local bulk dynamics, but from global infrared structures,
such as functional Berry holonomies and their associated
entanglement.

This viewpoint suggests that the appropriate arena for quantum
gravity is not the microscopic structure of spacetime, but the
infrared geometry perceived by external observers.

\begingroup
\let\clearpage\relax
\section*{ACKNOWLEDGMENTS}

The authors are grateful to R. Mackenzie and F. Mendez for useful discussions. This research was 
supported by DICYT (USACH), grant number 042531GR\_REG. The work of N.T.A is supported by Agnes Scott College.
\endgroup

\nocite{*} 
\bibliographystyle{JHEP}
\bibliography{ref.bib}

\end{document}